\RequirePackage{fix-cm}
\documentclass[twocolumn]{article}          
\usepackage{mathptmx}      
\usepackage{amssymb}
\usepackage{tikz}
\usepackage{pgf}
\usepackage{bbm}
\usetikzlibrary{er} 
\usetikzlibrary[er] 
\usetikzlibrary{snakes}   
\usepackage{graphicx}
\usepackage[]{xcolor}
\usepackage{mathtools}
\usepackage{soul}
\usepackage[makeroom]{cancel}
\usepackage{setspace}
\usepackage{lmodern}
\usepackage{scalefnt}
\usepackage{epigraph}
\usepackage{color}
\usepackage{dcolumn}
\usepackage{url}
\usepackage{changes}
\usepackage{authblk}

\usepackage{bm}
\newcommand{\miniscule}{\@setfontsize\miniscule{4}{5}}

\definechangesauthor[name={Christian}, color=red]{chris}
\definechangesauthor[name={Sue}, color=red]{sue}
\definechangesauthor[name={Hayato}, color=blue]{hayato}
\begin{document}
  
  \title{  
    Detecting Multiple Communities Using Quantum Annealing on the D-Wave System
  }
  
  %
  %
  \author[1]{Christian F. A. Negre}
  \author[2]{Hayato Ushijima-Mwesigwa}
  \author[3]{Susan M. Mniszewski}
  
  \affil[1]{Theoretical Division, Los Alamos National Laboratory, Los Alamos, NM 87545}
  \affil[2,3]{Computer, Computational, \& Statistical Sciences Division, Los Alamos National Laboratory, Los Alamos, NM 87545}
  
  \maketitle
  
  \begin{abstract}
    A very important problem in combinatorial optimization is partitioning a network into communities of densely connected nodes; where the connectivity between nodes inside a particular community is large compared to the connectivity between nodes belonging to different ones. This problem is known as community detection, and has become very important in various fields of science including chemistry, biology and social sciences. 
    The problem of community detection is a twofold problem that consists of determining the number of communities and, at the same time, finding those communities. 
    This drastically increases the solution space for heuristics to work on, compared to traditional graph partitioning problems.    
    In many of the scientific domains in which graphs are used, there is the need to have the ability to partition a graph into communities with the ``highest quality'' possible 
    since the presence of even small isolated communities can become crucial to explain a particular phenomenon. 
    We have explored community detection using the power of quantum annealers, and in particular the D-Wave 2X and 2000Q machines. 
    It turns out that the problem of detecting at most two communities naturally fits into the architecture of a quantum annealer with almost no need of reformulation. 
    This paper addresses a systematic study of detecting two or more communities in a network using a quantum annealer. 
    
  \end{abstract}
  
  \section{Introduction}
  \label{intro}
  
  The use of networks spans across many scientific domains. To showcase how broad this spectrum is, we highlight the following examples: molecules are chemical networks with atoms connected to each other \cite{niklasson2016,rivalta2012allosteric,Negre201810452}; living cells follow a communication pattern described by a network \cite{jeong2000}; and human interactions create different social networks \cite{ugander2011TheAO}. The study of networks has resurged in recent times due to the availability and capability of producing and storing data from a large number of applications. Advances in crystallography, for example, allows chemists to get atomistic representation of complex proteins; social media speeds up human communication; etc. \cite{fortunato2016}. 
  
  In all of the previous mentioned scientific domains in which graphs are used, there is the imminent need to have the ability to partition 
  a graph into communities with the ``highest quality'' possible. In other words we want the best community split to be able to recognize 
  particular features of the network such as the presence of small communities that could become crucial to explain a particular phenomenon \cite{rivalta2012allosteric}.
  Another 
  feature that needs to be properly revealed is the boundary
  between communities which could be crucial for classification.

\begin{figure}
    \begin{center}
      \vspace{-0pt}
      \includegraphics[width=0.50\textwidth]{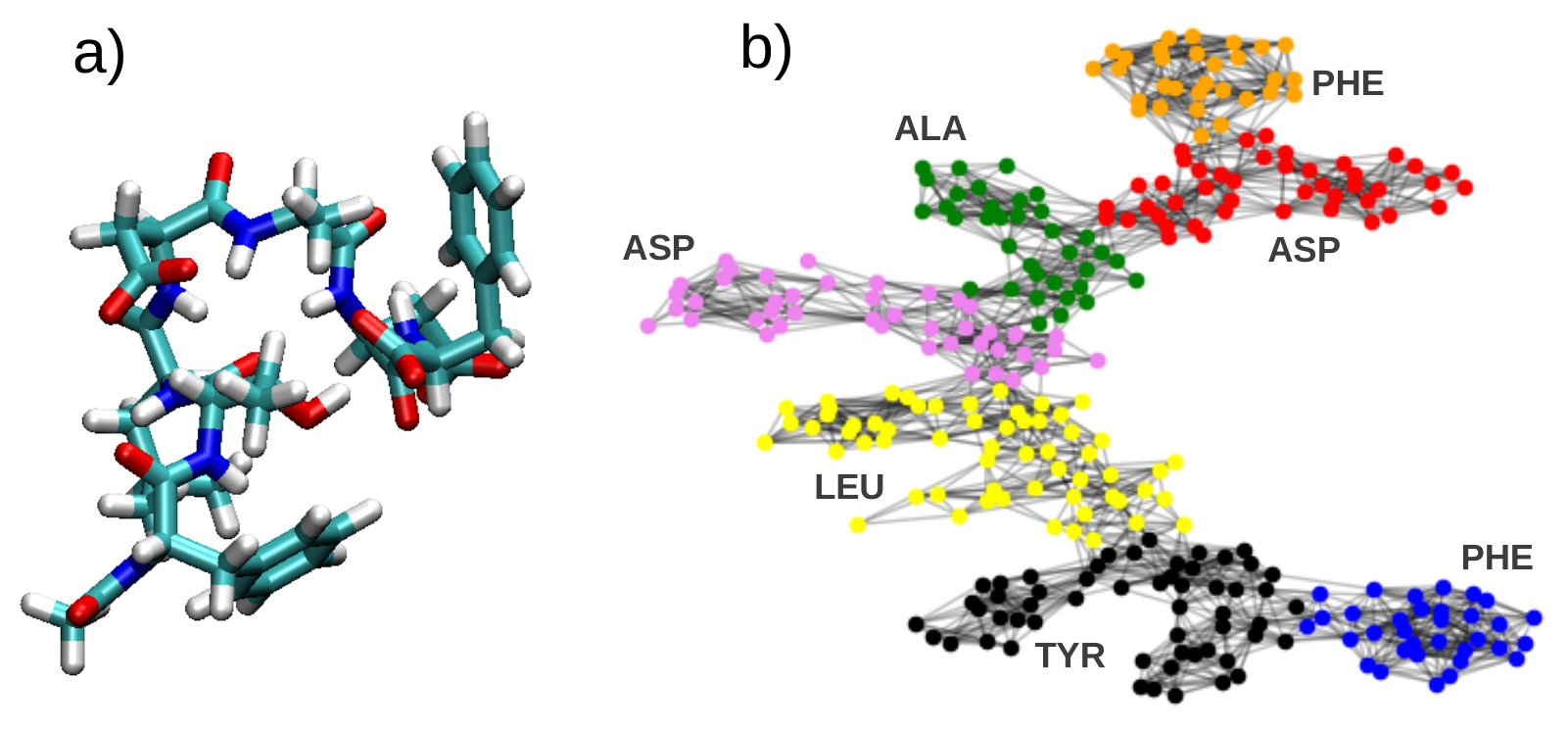}     
      \vspace{-0pt}
      \caption{\small a) Molecular representation of a small protein composed of seven amino acids (PHE, ASP, ALA, ASP, LEU, TYR and PHE). b) Representation of the orbital connectivity associated graph with nodes colored by the communities they belong to.}
      \vspace{-20pt}
      \label{alpha} 
    \end{center}
  \end{figure}

  Figure \ref{alpha}a shows the molecular structure of a small protein composed of seven amino acids \cite{ghale2017}. In Figure \ref{alpha}b, we show the result of partitioning a graph built out of the 
  the connectivity between orbitals of different atoms\footnote{This connectivity is computed out of the density matrix of the system, which is an object that accounts for the electronic structure.}. The graph is composed of 300 nodes/orbitals and 1794 edges/connections. 
  This community splitting was performed using the technique presented in this paper, rendering 7 communities with a modularity of 0.766. We can clearly see how, roughly, each amino acid that composes the protein is classified as belonging to a single community. This is far from being trivial. In order to recognize each amino acid component, a computational chemist will need to perform a careful inspection of the chemical structure of the protein, which could become a time-consuming task if the system is larger.

  Due to the large size of network datasets, many meaningful properties on networks are expensive to compute, especially when dealing with NP-hard graph problems. 
  Post Moore's era supercomputing has provided us an opportunity to explore new approaches for traditional graph algorithms on quantum computing architectures \cite{hayato2017}.
  These machines are attractive alternatives to traditional computers in helping to solve the aforementioned graph problems \cite{shaydulin2018}. 
  In a recent paper we have developed a formulation for graph partitioning to be performed on a quantum annealer \cite{hayato2017}. In this paper we present a formulation for performing community detection of multiple communities (two or more).
  
\begin{figure}[ht]
    \centering
   \includegraphics[width=9cm]{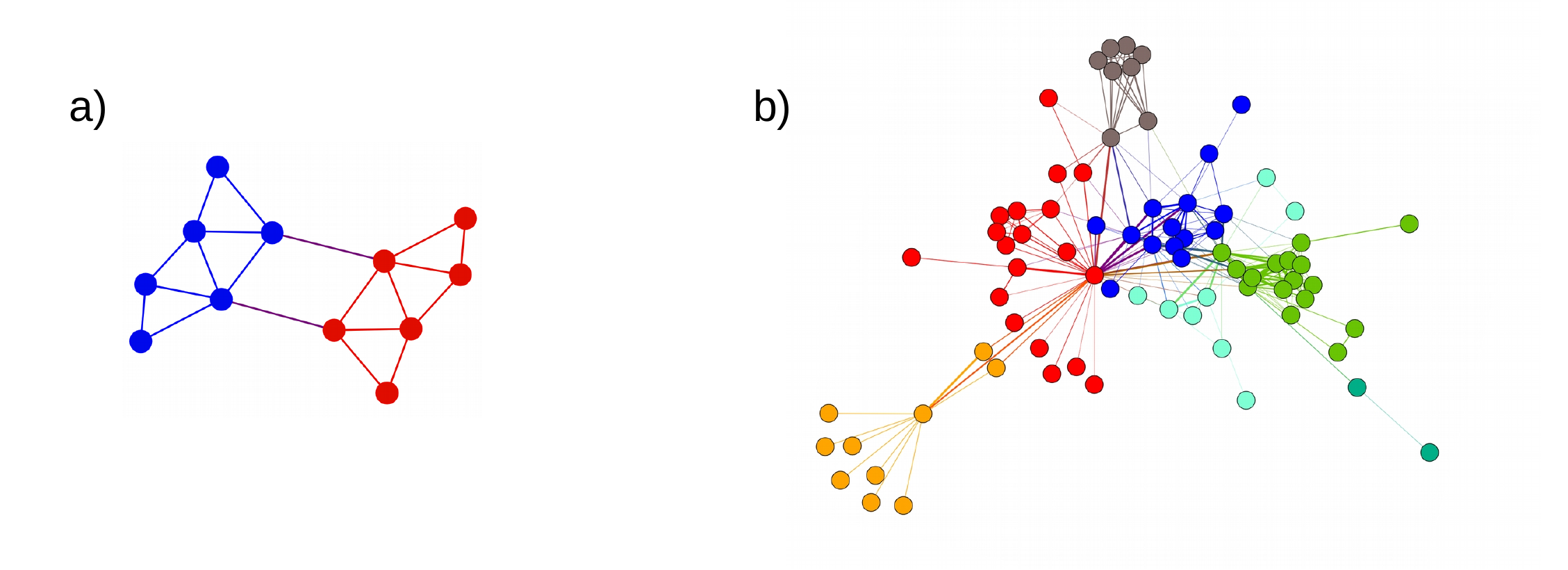}           
 \caption{ Graph partitioning (a) showing two parts of equal size (six nodes) depicted with red and blue colors; and (b), example of community detection showing 6 different communities of the Les Miserables coappearance social network.}
    \label{gp-clustering}
  \end{figure}

  The general concept of community structure was introduced for the first time by Girvan and Newman in 2002 \cite{girvan2002} to describe the general appearance (the skeleton) of a network. 
  A network can be divided into sets of nodes belonging to different communities (also called clusters). Nodes within any of these communities are highly connected (high intraconnectivity); whereas nodes in different communities are less connected (low interconnectivity). 
  This natural division of a graph into communities differs from the usual graph partitioning (GP) problem in that there are no restrictions on the size of the communities (see Figure \ref{gp-clustering}). Methods involving both community detection and graph partitioning are considered unsupervised machine learning techniques \cite{Geron2017}. 
  The total number of feasible solutions for k-community detection over a network of $n$ nodes is given by the Bell number $B_n = \frac{1}{e}\sum_{k=0}^{\infty}\frac{k^n}{k!}$ which has an upper bound of $\left(\frac{0.792n}{\ln(n+1)}\right)^n$ \cite{berend2010}. This implies an extensive search space for feasible solutions as compared to the regular balanced GP problem. 
  It also implies the need to combine heuristics and recursive optimization methods to search for the optimal community structure. 
  Several algorithms have been proposed based on greedy techniques \cite{newman2004a,blondel2008}, simulated annealing \cite{kirkpatrick1983,amaral2005}, genetic algorithm\cite{goldberg1989}, spectral optimization \cite{fortunato2010} and extreme optimization methods \cite{boettcher2001,duch2005}.
  
  The notion of ``more connected'' within a community is arbitrary and the existence of a unique definition is still under debate, however, the metric proposed by Girvan and Newman based on the modularity 
  has been well accepted in the field \cite{girvan2002,newman2004,newman2006}. 
  Other metrics and methods include $k$-means \cite{hartigan1979}, spectral \cite{shi2000}, hierarchical \cite{ward2963},
  modularity density \cite{newman2004}, min-max cut \cite{ding2001}, and normalized cut \cite{shi2000} among others. Moreover, Neukart, et al., implemented a quantum-assisted clustering method 
  on the D-Wave system \cite{Neukart2018}.
  The modularity metric allows to quantify the quality of a community structure by 
  comparing the connectivity of edges within communities with the connectivity of an equivalent network where edges would be placed randomly. 
  
  Maximizing the modularity with respect to the community partitions is a problem that has been largely studied and different methods has been proposed. 
  Despite all these efforts, modularity maximization still remains a field that needs further exploration. Two main problems have been identified. One 
  of them concerning the existence of a resolution limit in which a very small community could be missed from the identified community structure also containing large communities \cite{fortunato2007}. The second problem is the existence of several minima with the same modularity which prevents or makes it difficult to obtain the global minima \cite{good2010}. 
  These two problems are related to what we have previously loosely  defined as highly resolved community structure. 
  
  A method for further refining community structures has been recently introduced \cite{cafieri2014}. 
  The authors proposed an exact algorithm for bi-partitioning a network using split and merge movements 
  over communities of an existing partition. Using this approach the authors were able to further improve the modularity values and, as a consequence, 
  the quality of the community structure. We will use some of these highly refined results as a point of comparison with the ones obtained from the 
  method we are introducing. 
  
  Quantum computers of the annealer type, such as the D-Wave 2X and 2000Q, minimize the Ising objective function as follows making use of the quantum entanglement effect \cite{dwave}. The objective function can be written as follows: 
  \begin{equation}
    O(\textbf{h,J,s}) = \sum\limits_{i}h_{i}s_i + \sum\limits_{i<j}J_{ij}s_is_j
    \label{dwave}
  \end{equation}  
  where $s_{i} \in$ $\{-1,+1\}$ are magnetic spin variables where the result is encoded; $h_{i}$ and $J_{ij}$ are local magnetic fields and coupling strengths that encode the problem Hamiltonian. 
  At the hardware level, the D-Wave quantum computer is composed of qubits with sparse connectivity as a fixed sparse graph, known as a \emph{Chimera} graph. 
  During the annealing process each qubit can be in a ``superposition'' state (both a ``-1'' and a ``+1'' simultaneously). The superposition 
  lasts until an outside event causes it to collapse into either a ``-1'' or a ``+1'' state. 
  The result of the annealing process is a low-energy ground state $\textbf{s}$, consisting of an Ising spin for each qubit value $\in$ $\{-1,+1\}$. 
  This allows quantum computers to solve  NP-hard complex complex problems including optimization, machine learning and sampling problems.
  Maximization problems can also be solved by the D-Wave by using the negative of Equation \ref{dwave} as the objective function.
  The formulation where variables take values of either 0 or 1 is called the quadratic unconstrained binary optimization or QUBO formulation and it is an alternative representation that can easily be translated to or from the Ising model.
  An Ising model can become a QUBO through the transformation, $s = 2x - 1$.
  
Current D-Wave platforms have physical constraints such as limited precision, sparse connectivity, and number of available qubits.  Embedding is required to map a problem onto the hardware \emph{Chimera} graph prior to annealing. Purely quantum approaches are limited by the number of graph nodes/variables that can be represented on the hardware, 46 for the 2X and 64 for the 2000Q. Anything larger requires a hybrid classical-quantum approach.

  In this paper we describe the method for performing community detection based on the modularity metric using the D-Wave quantum annealer.
  We carefully derive the formulation of the problem as a QUBO.
  Results are compared with existing benchmarks using ``state of art'' tools.

  \section{Formulation}

  Let $G=(V,E)$ be a weighted graph with nodes $i$ in $V$ and edges $ij$ in $E$ such that the corresponding adjacency matrix $A$ is defined as follows: 
  \begin{equation}
    A_{ij} = \left\{\begin{array}{clrr}
      0, & \mathrm{if} \quad i=j  \\
      w_{ij}, &  \mathrm{if}  \quad  i\neq j
    \end{array}\right.
    \label{adj}
  \end{equation}
  with $w_{ij}$ being the weight of edge $ij$.
  We can then construct a modularity matrix $B$ as the difference between $A$ and a matrix constructed 
  as an outer product of the vector degree  $\textbf{g}$. Here, the node degree $g_i$ is defined as $g_i = \sum_j A_{ij}$. 
  Newman's expression for the modularity matrix $B$ can be written as follows: 
  \begin{equation}
    B = A - \frac{\textbf{g}\textbf{g}^T}{2m}
    \label{Q}
  \end{equation}
  Or equivalently:
  \begin{equation}
    B_{ij}=  A_{ij} - \frac{g_ig_j}{2m} =  A_{ij} - \frac{g_ig_j}{\sum_{l}g_{l}}
    \label{Qij}
  \end{equation}
  
  For partitioning the graph into at most two communities, if $s_i \in \{-1,1\}$, is a binary variable indicating which community node $i$ belongs to, then the modularity $Q$ for any given partition is given by 
  \begin{equation}
    Q = \frac{1}{2m} \sum_{i,j}B_{ij}\cdot \frac{s_i s_j + 1}{2}
    \label{2parts}
  \end{equation}

  Community detection requires maximizing the modularity $Q$.    
  The rows and columns of the matrix $B$ sum to zero, thus, we have $\sum_{i,j}B_{ij}\frac{s_i s_j + 1}{2}  = \sum_{i,j}B_{ij} \frac{s_i s_j}{2}$. In addition, the term $\frac{s_i s_j + 1}{2}$ can be viewed as a product of binary variables $x_i \in \{0,1\}$.   We can therefore write equation (\ref{2parts}) in matrix form as 
  
  \begin{equation*}
  Q = \frac{1}{4m}\textbf{s}^T B \textbf{s}
  \end{equation*}
  where $\textbf{s}$ is a column vector with entries $s_i$. 
  A transformation from a QUBO to Ising formulation can be given by
  \begin{equation*}
    \textbf{s}^T B\textbf{s} = 4\textbf{x}^T B\textbf{x} - 4\textbf{x}^T B \mathbbm{1} + \mathbbm{1} ^T B \mathbbm{1}
  \end{equation*}
  where $\textbf{x}$ is a vector such that $x_i\in \{0,1\}$, $\mathbbm{1}$ is a vector of all ones, and $\textbf{s} = 2 \textbf{x} - \mathbbm{1}$.
  However, since $B \mathbbm{1} = \textbf{0}$, we have 
  \begin{equation}
    \textbf{s}^T B\textbf{s} = 4\textbf{x}^T B\textbf{x} 
    \label{parts}
  \end{equation}
  Equation (\ref{parts}) essentially shows that both QUBO and Ising formulations are equivalent for modularity maximization. 
  Thus, the maximum modularity for  at most two communities is given by 
  \begin{equation}
    \max_{\textbf{s}}\left( \frac{1}{4m} \textbf{s}^T B \textbf{s} \right) \mbox{\quad or \quad } \max_{\textbf{x}}\left(\frac{1}{m}\textbf{x}^T B \textbf{x}\right) 
    \label{maxmod}
  \end{equation}
  which are clearly unconstrained quadratic optimization problems, suitable to be solved by quantum annealers. 
  
  Now, if we are interested in partitioning the graph into at most $k$ communities, the modularity for any given partition is given by
  \begin{equation}
    Q = \frac{1}{2m} \sum_{i,j}B_{ij}\cdot \delta(c_i, c_j)
    \label{deltaparts}
  \end{equation}
  where $1 \leq c_i \leq k$ is the community node $i$ belongs to and the function $\delta(c_i,c_j) = 1$, if $c_i = c_j$ or $0$ otherwise;
  with $i \neq j$ for $1\leq i$ and  $j \leq |V|$. 
  Equation \ref{deltaparts} poses the problem that the function $\delta$ is not necessarily a quadratic binary variable function.
  
  
  In our previous work, we demonstrated dividing a graph into two communities \cite{hayato2017}. Additionally,
  we introduced the concept of a logical super-node which allows the partition of graphs into more than two parts in an ``all at once'' $k$-concurrent fashion (see Figure \ref{superedge}). Each logical super-node represents a graph node. A super-edge represents a connection between graph nodes. The logical super-node utilizes a one-hot encoding to represent the selection of 1 out of $k$ communities \cite{state-encodings}.
  In this work, we will use the concept of the logical super-node and a proper formulation leading to a QUBO problem to partition the graph into more than two communities in a $k$-concurrent fashion.

  \tikzstyle{com1}=[ellipse,inner sep=0pt,minimum size= 0.5 cm,draw=black,fill=gray!40,thick]
  \tikzstyle{com2}=[ellipse,inner sep=0pt,minimum size= 0.5 cm,draw=black,fill=gray!40,thick]
  \tikzstyle{com3}=[ellipse,inner sep=0pt,minimum size= 0.5 cm,draw=black,fill=gray!40,thick]
  \tikzstyle{com4}=[ellipse,inner sep=0pt,minimum size= 0.5 cm,draw=black,fill=gray!40,thick]
  \tikzstyle{com5}=[ellipse,inner sep=0pt,minimum size= 0.5 cm,draw=black,fill=gray!40,thick]
  \tikzstyle{com6}=[ellipse,inner sep=0pt,minimum size= 0.5 cm,draw=black,fill=gray!40,thick]
  \tikzstyle{com7}=[ellipse,inner sep=0pt,minimum size= 0.5 cm,draw=black,fill=gray!40,thick]
  \tikzstyle{com8}=[ellipse,inner sep=0pt,minimum size= 0.5 cm,draw=black,fill=gray!40,thick]
  \tikzstyle{post}=[font=\tiny,-,very thick, fill=white]
  \begin{figure}[h]
    \centering
    \begin{tikzpicture}[scale=0.70, inner sep=1.5mm, node distance=4cm]     
      \draw[dashed,thick] (0,0) circle [radius=1.5];
      \node at ( 0, -2 ) (tagj) {Super-node I};      
      \node at ( 0.0, 1.0 ) (com1) [com1] {$x_{i,1}$};
      \node at ( 1.0, 0.0 ) (com2) [com2] {$x_{i,2}$};
      \node at ( 0.0, -1.0 ) (com3) [com3] {$x_{i,3}$};
      \node at (-1.0, 0.0 ) (com4) [com4] {$x_{i,4}$};
      \node at (5.0, 3.0 ) (com5) [com5] {$x_{j,1}$};
      \node at (6.0, 2.0 ) (com6) [com6] {$x_{j,2}$};
      \node at ( 5.0,1.0 ) (com7) [com7] {$x_{j,3}$};
      \node at ( 4.0,2.0 ) (com8) [com8] {$x_{j,4}$};
      \foreach \x in {2} {
        \draw [line width={\x* 1.00},-,color=blue] (com1) -- (com2) ;
        \draw [line width={\x* 1.00},-,color=blue] (com1) -- (com3) ;
        \draw [line width={\x* 1.00},-,color=blue] (com1) -- (com4) ;
        \draw [line width={\x* 1.00},-] (com1) -- (com5) node[pos=0.4,above]{Q$_{I,J}$} ;
        \draw [line width={\x* 1.00},-,color=blue] (com2) -- (com3) ;
        \draw [line width={\x* 1.00},-,color=blue] (com2) -- (com4) ;
        \draw [line width={\x* 1.00},-] (com2) -- (com6) ;
        \draw [line width={\x* 1.00},-,color=blue] (com3) -- (com4) node[pos=1.4,xshift=-0.09cm,below]{Q$^{I}_{3,4}$};
        \draw [line width={\x* 1.00},-] (com3) -- (com7) ;
        \draw [line width={\x* 1.00},-] (com4) -- (com8) ;
        \draw [line width={\x* 1.00},-,color=blue] (com5) -- (com6) ;
        \draw [line width={\x* 1.00},-,color=blue] (com5) -- (com7) ;
        \draw [line width={\x* 1.00},-,color=blue] (com5) -- (com8) ;
        \draw [line width={\x* 1.00},-,color=blue] (com6) -- (com7) ;
        \draw [line width={\x* 1.00},-,color=blue] (com6) -- (com8) ;
        \draw [line width={\x* 1.00},-,color=blue] (com7) -- (com8) ;
      }
      \draw[dashed,thick] (5,2) circle [radius=1.5];
      \node at ( 5, 0 ) (tagj) {Super-node J};
    \end{tikzpicture}    
    \caption{An example of the super-node concept used in $k$-concurrent community detection is shown for partitioning into 4 parts. Two super-nodes $I$ and $J$ consisting of four subnodes ($x_{i/j,k}$) where each are connected by a super-edge Q$_{I,J}$. Internal edges \textcolor{blue}{Q$^{I/J}_{l,m}$} where $l,m \in \{1-4\}$ are set to enforce the selection of only one subnode to be equal to ``1'' after the annealing. The super-edge Q$_{I,J}$ is shown with connections between corresponding subnodes.}
    \label{superedge}
  \end{figure}
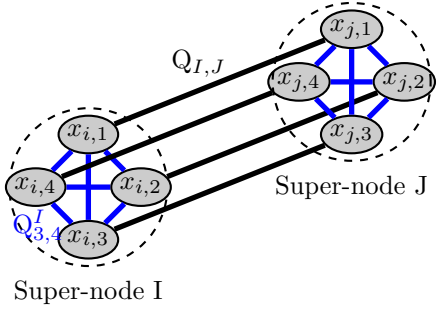

  In this section, we will generalize the modularity matrix formulation for an arbitrary number of communities. 
  For this, let $G = (N,E)$ be a graph for which the final result of the community split is a set composed of communities $C_i$ with the following property: $N = \left\{C_1 \cup C_2 \cup \ldots\right\}$.
  For a given community, $C$, the value 
  \begin{equation*}
    Q_C = \frac{1}{2m} \sum_{(i,j)\in C}\big(A_{ij} - \frac{g_ig_j}{2m}\big)
  \end{equation*}
  computes the modularity of community $C$ within the QUBO formulation.
  
  Suppose that now we want to partition $G$ into at most $k$ communities. Let $x_{i,j}$ be the decision variables such that
  \begin{equation*}
    x_{i,j} = \begin{cases}
      1 & \text{if node $i$ is in community $j$}\\
      0 & \text{otherwise}.
    \end{cases}
  \end{equation*}
  Strictly speaking, the set $\left\{x_{i,1}, \ldots x_{i,k}\right\}$ is a logical super-node as depicted in Figure \ref{superedge}. 
  Since each node must be in exactly one community, the following constraint needs to be fulfilled.
  \begin{equation}
    \sum_{j=1}^{k}x_{i,j} = 1
    \label{constraint}
  \end{equation}
  Let
  \begin{equation*}
    \textbf{x}_j = \begin{bmatrix}
      x_{1,j}  \\
      x_{2,j} \\
      \vdots \\
      x_{n,j}
    \end{bmatrix},
  \end{equation*}
  be the vector state, then:
  \begin{align}
    \begin{array}{ll@{}ll}
      Q(\textbf{x}) = \max_{\textbf{x}}(\sum_{j=1}^{k} \textbf{x}_j^TB\textbf{x}_j) 
    \end{array}
    \label{k_communities}
  \end{align}
  subject to  $\displaystyle\sum_{j = 1}^{k}x_{i,j} = 1 \, \text{, for } \, i = 1, \dots,n$
  with $ x_{i,j} \in \{0,1\}$ for $i=1 ,..., n$  and $j = 1, \dots, k$.
  A corresponding relaxation is given by
  \begin{align}
    \small
    \begin{split}
      \max_{\textbf{x}}\Big(\sum_{j=1}^{k} \textbf{x}_j^TB\textbf{x}_j 
      +\sum_{i = 1}^{n} \gamma_i(\sum_{j = 1}^{k}x_{i,j} - 1)^2\Big) 
    \end{split}
    \label{m_communities_relax}
  \end{align}    
  with $x_{i,j} \in \{0,1\},\text{for } i=1 ,..., n$ and $ j = 1, \dots, k$. $\gamma_i$ are the corresponding relaxation coefficients.

  The derivation that follows has already been shown in reference \cite{hayato2017}. We will give a summary for completeness here. We start by considering 
  the following equality:
  \begin{align*}
    ( \sum_{j=1}^{k}x_{i,j} - 1)^2 &=(\sum_{j=1}^{k}x_{i,j} )^2 -2\sum_{j=1}^{k}x_{i,j}  + 1
  \end{align*}
  \noindent Let now $\textbf{Z}_i$ be the $\mathcal{N}\times \mathcal{N}$ zero matrix (with $\mathcal{N} = k \times n$) whose $j$th diagonal element is 1 if and only if $j \equiv i\ (\textrm{mod}\ n)$. For example, in $\textbf{Z}_1$ every $1^{\text{st}}, (n+1)^{\text{th}}, (2n+1)^{\text{th}},\dots, ((k-1)n+1)^{\text{th}}$ diagonal element is 1 and has zero everywhere else. Then
  \begin{align*}
    \big(\sum_{j=1}^{k}x_{i,j}\big)^2 & = \vec{X}^T\vec{Z}_i \mathbbm{1}_{\mathcal{N}\times\mathcal{N}}\vec{Z}_i\vec{X}
  \end{align*}
  where $\vec{X}^T$ is the vector $(x_{1,1},\ldots,x_{n,1},\ldots,x_{1,k},\ldots, x_{n,k})$, and 
  \begin{align*}
    \sum_{j=1}^{k}x_{i,j} &= \mathbbm{1}_{\mathcal{N}}^T\textbf{Z}_i\textbf{X}.
  \end{align*}
  Hence,
  \begin{align*}
    ( \sum_{j=1}^{k}x_{i,j} - 1)^2 &= \textbf{X}^T\textbf{Z}_i \mathbbm{1}_{\mathcal{N}\times\mathcal{N}}\textbf{Z}_i\textbf{X} - 2 \mathbbm{1}_{\mathcal{N}}^T\textbf{Z}_i\textbf{X}+ 1
  \end{align*}
  and
  \begin{align*}
    \begin{split}
      \sum_{i=1}^{n} \gamma_i( \sum_{j=1}^{k}x_{i,j} - 1)^2 &= \sum_{i=1}^{n} \gamma_i \Big(\textbf{X}^T\textbf{Z}_i \mathbbm{1}_{\mathcal{N}\times\mathcal{N}}\textbf{Z}_i\textbf{X} \\
      &-2 \mathbbm{1}_{\mathcal{N}}^T\textbf{Z}_i\textbf{X}+ 1\Big)\\
      &= \textbf{X}^T\sum_{i=1}^{n} \gamma_i \Big(\textbf{Z}_i \mathbbm{1}_{\mathcal{N}\times\mathcal{N}}\textbf{Z}_i\Big)\textbf{X} \\
      &- 2\sum_{i=1}^{n} \gamma_i \mathbbm{1}_{\mathcal{N}}^T\textbf{Z}_i\textbf{X}+ \sum_{i=1}^{n} \gamma_i.
    \end{split}
  \end{align*}
  Let $ D_{\gamma} $ be a diagonal matrix such that 
  \begin{equation*}
    D_{\gamma} = \textrm{diag}(\gamma_1, \dots, \gamma_n)
  \end{equation*}
  and
  $\textbf{B}_{\Gamma}$ be a block matrix with $k\times k$ blocks, where each block is equal to $D_{\gamma}$, then 
  \begin{equation*}
    \sum_{i=1}^{n} \gamma_i \textbf{Z}_i \mathbbm{1}_{\mathcal{N}\times\mathcal{N}}\textbf{Z}_i = \textbf{B}_{\Gamma}
  \end{equation*}
  and 
  \begin{equation*}
    \sum_{i=1}^{n}\gamma_i\mathbbm{1}_{\mathcal{N}}^T\textbf{Z}_i\textbf{X} = \Gamma^T\textbf{X}.
  \end{equation*}
  where we have defined $\Gamma^T = \sum_{i=1}^{n}\gamma_i\mathbbm{1}_{\mathcal{N}}^T\textbf{Z}_i$. 
  So,
  \begin{align*}
    \sum_{i=1}^{n} \gamma_i( \sum_{j=1}^{k}x_{i,j} - 1)^2     &= \textbf{X}^T\textbf{B}_{\Gamma}\textbf{X} - 2\Gamma^T\textbf{X}+ \sum_{i=1}^{n} \gamma_i. 
  \end{align*}

  \noindent Therefore, we have
  \begin{align}
    \begin{array}{ll@{}ll}
      & \max_{\textbf{x}}( \textbf{X}^T(\beta\mathcal{B} + \textbf{B}_{\Gamma})\textbf{X} - 2\Gamma^T\textbf{X}) &\\
      \\
      \text{with} &   x_{i,j} \in \{0,1\}, \quad   i=1 ,..., n,  \ j = 1, \dots, k
    \end{array}
    \label{k_communities_matrix}
  \end{align}  
  where $\mathcal{B}$ is a block diagonal matrix with the modularity matrix $B$ on the diagonal.
  \section{Results and discussion}

  
We used the python NetworkX tools \cite{aric2008} for pre- and post-processing of the example graphs. The size of these graphs in most cases is larger than the number of nodes or variables that can be embedded on the D-Wave 2X and 2000Q. Therefore, we have used the hybrid classical-quantum tool, 
\emph{qbsolv}, developed by D-Wave \cite{booth2017}. The \emph{qbsolv} software takes the full problem in QUBO format as input and makes multiple calls to the D-Wave to solve subQUBOs for global minimization, followed by tabu search for local minimization.
It can be called directly through the D-Wave Ocean application programming interface (API) or from the command line. Resulting bitstrings of zeros and ones are translated based on the optimization problem's representation.
  
\begin{figure}[ht]
    \centering
    \includegraphics[width=7cm]{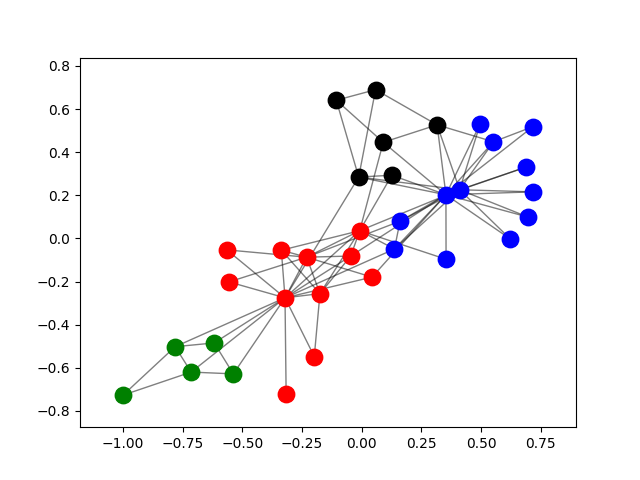}           
    \caption{Zachary karate club network showing the different communities find with the D-Wave quantum annealer.    
    }
    \label{karate}
  \end{figure}

  The quality of a community structure is determined by evaluating the modularity metric. This varies from 0 to 1, with a larger value being preferable.
  We have used the Zachary (karate club) graph to compare the results of the modularity metric obtained with other methods.
  The aforementioned is a social network of friendships between 34 members of a karate club at a US university in the 1970s. With only 34 nodes and 78 edges, this  
  graph is considered an archetypal social network that has been extensively used to benchmark graph algorithms \cite{zachary2018}. 
  From Table \ref{methods} we observe that modularity is similar across all methods, however the quantum annealer approach results in the best value based on the community structure. 
  This value is identical to the record value obtained by Blondel et al. showing four communities with a modularity of 0.41979 \cite{blondel2008}. A representation of the community structure obtained by this method can be seen in Figure \ref{karate}.
  
  One of the benefits of using quantum annealing is that there
  is no need to proceed recursively. The total time for annealing is about 20 $\mu$s for the case of the D-Wave.

  \begin{table}[h]
    \begin{center}
      \begin{tabular}{l|c}
        Method& Modularity\\ \hline
        GN & 0.401\\
        CNM & 0.381\\
        DA & 0.419\\
        Newman &0.419\\ 
        QA & 0.420 \\ \hline
      \end{tabular}
      \caption{Values of modularity are shown for the Zachary network computed with different algorithms: Girvan and Newman (GN) \cite{girvan2002}; Clauset et al. (CNM) \cite{clauset2004};  Duch and Arenas (DA) \cite{duch2005}; and the D-Wave quantum annealer (QA).}
      \label{methods}
    \end{center}
  \end{table}


  \begin{figure}[ht]
    \centering
    \includegraphics[width=5cm]{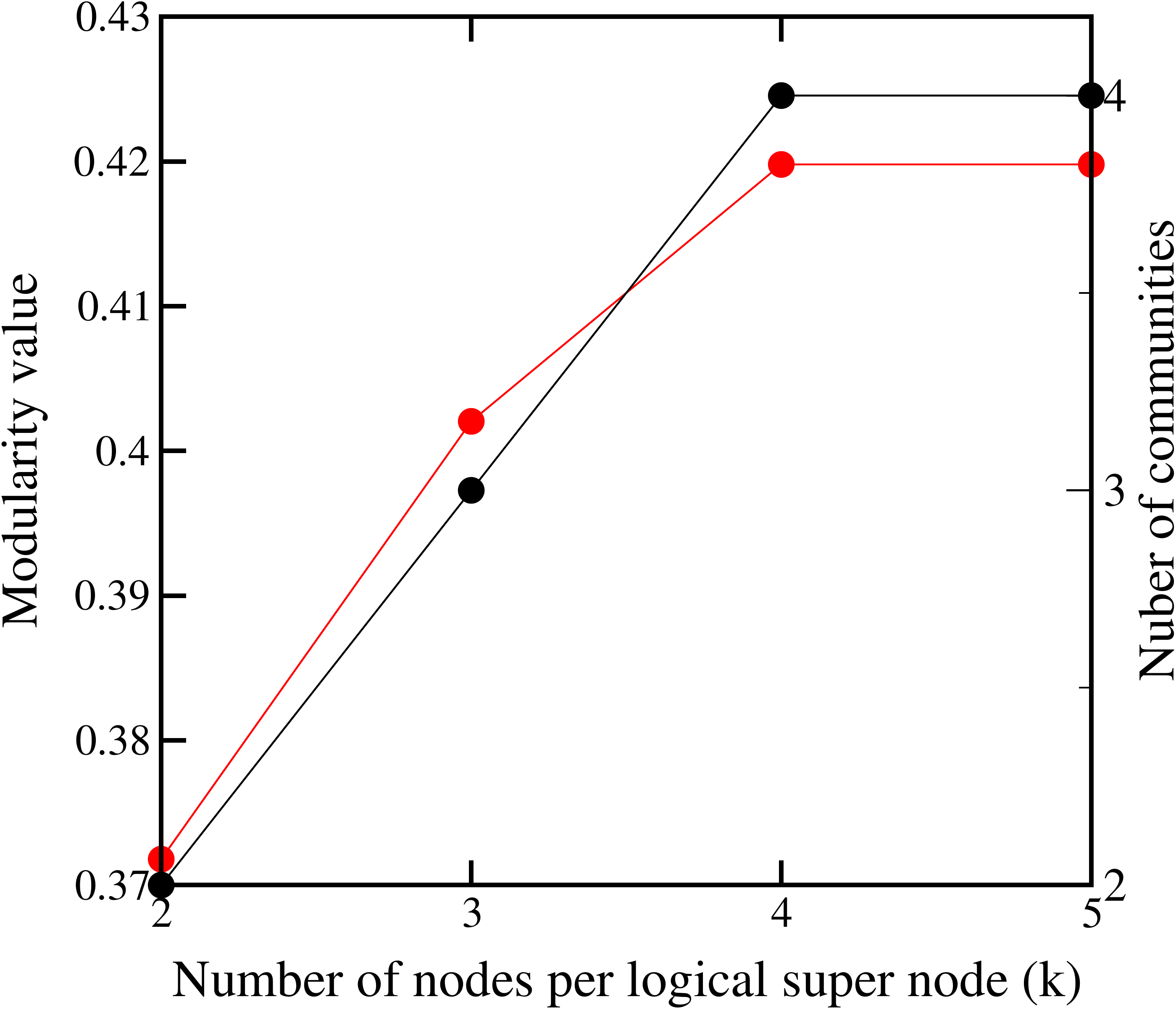}           
    \caption{ Convergence of community detection for the Zachary network with increasing number of nodes (or variables) per logical super-node. The black dots 
      show the number of communities that are obtained and the red dots show the modularity values.       
    }
    \label{convergence}
  \end{figure}  
  
  In Figure \ref{convergence} we can see how the numbers of communities and the modularity reach a maximum when the number of nodes (or variables) per logical super-node increases. 
  Modularity and number of communities reach values of 0.42 and 4 respectively. Instead of increasing the number of nodes per super-node iteratively, one could directly use a large number such as $k=n$ and 
  get the same result. 
  Currently, the D-Wave machines are limited in the problem size as number of variables that can be embedded and run on the \emph{Chimera} graph architecture. The 2X is limited to 46 variables, while the 2000Q can run problems up to 64.  
The sparse connectivity of the \emph{chimera} graph requires that some variables will be represented by chains of qubits \cite{lanting2014}. This can quickly use up the available qubits on the 2X (up to 1152) or the 2000Q (up to 2048).
  %
  The penalty constant $\gamma$ that is used to constrain each graph node to be in only one community can vary depending on the graph. Tuning is required for each new network problem.
  
  %

  We have computed the community structure for the following set of benchmark social networks: a coappearance network of characters in the novel Les Miserables (LesMiserables) \cite{knuth1993}; 
  an undirected social network of frequent associations between 62 dolphins in a community living off Doubtful Sound, New Zealand (Dolphins) \cite{lusseau2003};
  a book co-purchasing network where nodes represent books about US politics and edges represent frequent co-purchasing of books (PoliticalBooks) \cite{newman2006};
  a collaboration network of Jazz musicians (Jazz) \cite{kaur2016};
  and a metabolic network of the nematode C. elegans (Elegans) \cite{duch2005}. Results are shown in Table \ref{graphs}.
  
  \begin{table}[htbp]
    \caption{The results of $k$-concurrent community detection on benchmark graphs. $N$, $E$, $N_{com}$, and $Mod.$ are the number of nodes, number of edges, number of communities and modularity respectively.}
    \begin{center}
      \begin{tabular}{l|r|r|r|r}
        \hline
        & \multicolumn{1}{l|}{$N$} & \multicolumn{1}{l|}{$E$} & \multicolumn{1}{l|}{$N_{com}$} & \multicolumn{1}{l}{$Mod.$} \\ \hline
        Zachary & 34 & 78 & 4 & 0.41979 \\ 
        Dolphins & 62 & 159 & 5 & 0.52852 \\
        LesMiserables & 77 & 254 & 6 & 0.55861 \\ 
        PoliticalBooks & 105 & 441 & 4 & 0.52555 \\ 
        Jazz  & 198 & 2742 & 3 & 0.44447 \\ 
        Elegans  & 453 & 2040 & 5 & 0.41728 \\ \hline
      \end{tabular}    
    \end{center}
    \label{graphs}
  \end{table}

  Results for the Zachary and Dolphins benchmarks match the highly refined values reported in \cite{cafieri2014} which, to our knowledge are the ``best'' in the literature ever obtained. 
  The quantum annealer performs slightly better for the Dolphins network, when compared with non-refined results in the literature \cite{Fu2017}.
  Our results on LesMiserables and PoliticalBooks benchmarks are similar in comparizon to the refined values of 0.56001 and 0.52724 respectively \cite{Fu2017}. 
  On the other hand, results for the Jazz and Elegans benchmarks are similar in comparizon, but lower than the values reported in the literature \cite{duch2005}. 
  
  The score performed by the annealer reflects the size of the machine architecture. A small graph results in a clean embedding on the \emph{Chimera} graph, 
  with small chains to compensate for missing connections. As we said before, when graphs are larger we start having longer chains which can hinder the annealing performance \cite{lanting2014}. 
  For even larger graphs, hybrid classical-quantum approaches such as \emph{qbsolv} are used for orchestrating the use of the D-Wave solver on subproblems.
  Lower performance using \emph{qbsolv} may come from the mixing of classical and quantum processing.
 The quantum state initially created is no longer coming from the full Hamiltonian, but several pieces of it and a full pseudo solution is reconstructed. 
 We strongly believe that having more qubits available
and an increased connectivity will improve the performance of this technique. 
  
  A particular property of the modularity matrix is that it can be thresholded with almost no effect on the quality of the community structure. 
  This statement has a significant impact since it is saying that we can save qubits and reduce the size of the problem to be embedded. 
  Here we have tested this hypothesis by using the Zachary graph and thresholding the modularity matrix weights before solving the problem for community detection. From Table \ref{thresh} we can see that we can remove up to 40\% of the edges (by thresholding) and still get the same community structure with the same modularity value. 

  \begin{table}[htbp]
    \caption{Results for $k$-concurrent community detection with thresholding on the Zachary benchmark are shown. $Thres.$, $E$, $N_{com}$, and $Mod.$ are the threshold values, number of edges, number of communities and modularity respectively.}
    \begin{center}
      \begin{tabular}{c|c|c|c}
        \hline
        $Thres.$ & $E$ & $N_{com}$ & $Mod.$ \\ \hline
        0.00 & 561 & 4 & 0.41978 \\ 
        0.02 & 544 & 4 & 0.41978 \\ 
        0.05 & 411 & 4 & 0.41978 \\ 
        0.06 & 334 & 4 & 0.41978 \\ 
        0.07 & 300 & 4 & 0.41510 \\ \hline 
        0.08 & 244 & 3 & 0.39907 \\ 
        0.10 & 227 & 3 & 0.39907 \\ \hline 
        0.15 & 169 & 2 & 0.37179 \\
        0.25 & 110 & 2 & 0.37179 \\ \hline
      \end{tabular}  
    \end{center}
    \label{thresh}
  \end{table}

  \section{Conclusion}
  
  In this paper we have demonstrated the ability of a quantum computer and in particular a quantum annealer to solve community detection as an optimization problem.
  From the analysis of the results we observe that the quantum computer can render a highly optimized community structure. In the case of the Zachary graph, 
  we reach the actual record value. For other other benchmark graphs, including the larger graphs such as Jazz and Elegans, the quality of the community structure is comparable to ``state of the art'' results. 
  
  One of the most notable observations is that by using this quantum annealing technique with the $k$-concurrent method, we obtain the community structure ``all at once'' within the annealing time.
  There is no need to implement an iterative process as is the case for heuristic methods run on classical computers. The quantum annealer naturally finds the optimal solution. Limitations such as a reduced number of available qubits and sparse connectivity can contribute to the quality of the results. Generally speaking we can reaffirm the idea that quantum annealers have the non-trivial capacity to find solutions in a virtually instantaneous way.
  
  \section{Acknowledgements}  
  We acknowledge D-Wave Systems for their useful tutorials and use of the Burnaby D-Wave 2000Q machine. NNSA's Advanced Simulation and Computing (ASC) program at Los Alamos National Laboratory (LANL) for use of their Ising D-Wave 2X quantum computing resource. 
  This research has been funded by the Los Alamos National Laboratory (LANL) Information Science and Technology Institute (ISTI), Laboratory Directed Research and Development (LDRD), and NNSA's Advanced Simulation and Computing (ASC) program. Assigned: Los Alamos Unclassified Report LA-UR-18-30760. 
  Los Alamos National Laboratory is operated by Triad National Security, LLC, for the National Nuclear Security Administration of U.S. Department of Energy (Contract No. 89233218NCA000001).    
  
   \bibliographystyle{unsrt}
  \bibliography{biblio.bib}   

\end{document}